\documentclass[11pt]{article} 
\usepackage{amssymb}
\usepackage{amsmath}
\usepackage{natbib}
\usepackage{graphicx}
\usepackage{algorithmic}
\usepackage{algorithm}
\usepackage{subfigure}
\usepackage{amsthm}
\usepackage{epstopdf}

\usepackage[margin=3cm]{geometry}

\begin{document}

\newcommand{\vect}[1]{\boldsymbol{#1}}

\setlength{\parindent}{0pc}
\setlength{\parskip}{1ex}


\title{ABC and Indirect Inference} 
\author{Christopher C Drovandi \\ \\School of Mathematical Sciences, Queensland University of Technology, Australia}
\maketitle

ABSTRACT

Indirect inference (II) is a classical likelihood-free approach that pre-dates the main developments of ABC and relies on simulation from a parametric model of interest to determine point estimates of the parameters.  It is not surprising then that some likelihood-free Bayesian approaches have harnessed the II literature.  This chapter provides an introduction to II and details the connections between ABC and II.  A particular focus is placed on the use of an auxiliary model with a tractable likelihood function, an approach commonly adopted in the II literature, to facilitate likelihood-free Bayesian inferences.

\section{Introduction}

Indirect inference (II) is a classical method for estimating the parameter of a complex model when the likelihood is unavailable or too expensive to evaluate.  The idea was popularised several years prior to the main developments in ABC by \citet{Gourieroux1993,Smith1993}, where interest was in calibrating complex time series models used in financial applications.  The II method became a very popular approach in the econometrics literature (e.g.\ \citet{Smith1993,Monfardini1998,Dridi2007}) in a similar way to the ubiquitous application of ABC to models in population genetics.  However, the articles by \citet{Jiang2004} and \citet{Heggland2004} have allowed the II approach to be known and appreciated by the wider statistical community.

In its full generality, the II approach can be viewed as a classical method to estimate the parameter of a statistical model on the basis of a so-called indirect or auxiliary summary of the observed data \citep{Jiang2004}.  A special case of II is the simulated method of moments \citep{McFadden1989}, where the auxiliary statistic is a set of sample moments.  In this spirit, the traditional ABC method may be viewed as a Bayesian version of II, where prior information about the parameter may be incorporated and updated using the information about the parameter contained in the summary statistic.  However, much of the II literature has concentrated on developing the summary statistic from an alternative parametric auxiliary model that is analytically and/or computationally more tractable.  The major focus of this chapter is on approximate Bayesian methods that harness such an auxiliary model. These are referred to as parametric Bayesian indirect inference (pBII) methods by \citet{DrovandiBII2014}.

One such approach, referred to as ABC II, uses either the parameter estimate or the score of the auxiliary model as a summary statistic for ABC.  When the auxiliary parameter is used, the ABC discrepancy function (distance between observed and simulated data) may be based on a direct comparison of the auxiliary parameter estimates (ABC IP where P denotes parameter \citep{DrovandiEtAl2011}) or indirectly via the auxiliary log-likelihood (ABC IL where L denotes likelihood \citep{Gleim}).  Alternatively, a discrepancy function can be formulated by comparing auxiliary scores (ABC IS where S denotes score \citep{Gleim}).   Another approach, which differs substantially from the ABC II methods in terms of its theoretical underpinnings, uses the likelihood of the auxiliary model as a replacement to the intractable likelihood of the specified (or generative) model provided that a mapping has been estimated between the generative and auxiliary parameters \citep{Reeves2005,Gallant2009}.  This method has been referred to as parametric Bayesian indirect likelihood (pBIL) by \citet{DrovandiBII2014}.  These methods will be discussed in greater detail in this chapter.

This chapter begins with a tutorial and a summary of the main developments of II in Section \ref{sec:II}.  This is followed by a review of ABC II methods in Section \ref{sec:ABCII}.  Section \ref{sec:BIL} describes the pBIL approach to approximate Bayesian inference using ideas from II.  Other connections between ABC and II are provided in Section \ref{sec:Further} for further reading.  The methods are illustrated on an infectious disease model and a spatial extremes application in Section \ref{sec:Examples}.  The chapter is summarised in Section \ref{sec:Discussion}, which also discusses some possible future directions for utilising or building upon the current Bayesian II literature.

\section{Indirect Inference} \label{sec:II}

The purpose of this section is to give an overview of the II developments in the classical framework.  It is not our intention to provide a comprehensive review of the II literature, but rather to provide a tutorial on the method and to summarise the main contributions.  

Assume that there is available a parametric auxiliary model with a corresponding likelihood $p_A(y_{obs}|\phi)$, where $\phi$ is the parameter of this model.  The auxiliary model could be chosen to be a simplified version of the model of interest (the so-called generative model here) or simply a data-analytic model designed to capture the essential features of the data.  The majority of the literature on financial time series applications has considered the former; for example, \citet{Monfardini1998} considers autoregressive moving average auxiliary models to estimate the parameters of a stochastic volatility model.  In contrast, in an ABC application, \citet{DrovandiEtAl2011} use a regression model that has no connection to the assumed mechanistic model, in order to summarise the data.   

The main objective of II is to determine the relationship between the generative parameter, $\theta$, and the auxiliary parameter, $\phi$.  This relationship, denoted here as $\phi(\theta)$, is often referred to as the mapping or binding function in the II literature.  A mathematical definition for this function is provided below.  If the binding function is known and injective (one-to-one), then the II estimate based on observed data $y_{obs}$ is $\theta_{obs} = \theta(\phi_{obs})$, where $\theta(\cdot)$ is the inverse mapping and $\phi_{obs}$ is the estimate obtained when fitting the auxiliary model to the data.  The II approach essentially answers the question, what is the value of $\theta$ that could have produced the auxiliary estimate $\phi_{obs}$?    In this sense, the II approach acts as a correction method for assuming the wrong model.  

Unfortunately the binding function, $\phi(\theta)$, is generally unknown, but it can be estimated via simulation.  Firstly, using a similar notation and explanation in \citet{Heggland2004}, define an estimating function, $Q(y_{obs};\phi)$, for the auxiliary model.  This could be, for example, the log-likelihood function of the auxiliary model.  Before the II process begins, the auxiliary model is fitted to the observed data
\begin{align*}
\phi_{obs} &= \arg \max_{\phi}Q(y_{obs};\phi).
\end{align*}
For a particular value of $\theta$, the process involves simulating $n$ independent and identically distributed (iid) datasets from the generative model, $y_{1:n} = (y_1,\ldots,y_n)$.  Each replicate dataset $y_i$, $i = 1,\ldots,n$, has the same dimension as the observed data $y_{obs}$.  Then, the auxiliary model is fitted to this simulated data to recover an estimate of the binding function, $\phi_n(\theta)$:
\begin{align*}
\phi_n(\theta) &= \arg \max_{\phi}Q(y_{1:n};\phi),
\end{align*}
where $Q(y_{1:n};\phi) = \sum_{i=1}^n Q(y_{i};\phi)$.  The binding function is defined as $\phi(\theta) = \lim_{n \rightarrow \infty} \phi_n(\theta)$.  There is an alternative representation of $\phi_n(\theta)$.  Define the estimated auxiliary parameter based on the $i$th simulated dataset as
\begin{align*}
\phi(\theta,y_i) &= \arg \max_{\phi}Q(y_i;\phi).
\end{align*}
Then we obtain $\phi_n(\theta)$ via
\begin{align*}
\phi_n(\theta) &= \frac{1}{n}\sum_{i=1}^n\phi(\theta,y_i).
\end{align*}
When this formulation for $\phi_n(\theta)$ is used, the definition of the binding function can also be represented as $E[\phi(\theta,y)]$ with respect to $f(y|\theta)$.   The II procedure then involves solving an optimisation problem to find the $\theta$ that generates a $\phi_n(\theta)$ closest to $\phi_{obs}$:
\begin{align}
\theta_{obs,n} &= \arg \min_{\theta}\{(\phi_n(\theta) - \phi_{obs})^\top W(\phi_n(\theta) - \phi_{obs})\}, \label{eq:II}
\end{align}
\citep{Gourieroux1993} where the superscript $\top$ denotes transpose and $\theta_{obs,n}$ is the II estimator, which will depend on $n$.  \citet{Gourieroux1993} show that the asymptotic properties of this estimator is the same regardless of what form is used for $\phi_n(\theta)$.   Note that equation \eqref{eq:II} assumes that the auxiliary parameter estimates $\phi_n(\theta)$ and $\phi_{obs}$ are unique.  The II estimator should have a lower variance by increasing $n$ but this will add to the computational cost.  Note that the above estimator will depend on the weighting matrix $W$, which needs to be positive definite.     This matrix allows for an efficient comparison when the different components of the auxiliary estimator have different variances and where there is correlation amongst the components.  One simple choice for the matrix $W$ is the identity matrix.  Another choice is the observed information matrix $J(\phi_{obs})$, which can be used to approximate the inverse of the asymptotic variance of $\phi_{obs}$.  Discussion on more optimal choices of the weighting matrix (in the sense of minimising the asymptotic variance of the indirect inference estimator) is provided in \citet{Gourieroux1993} and \citet{Monfardini1998}, for example. 

An alternative approach proposed in \citet{Smith1993} is to set the II estimator as the one that maximises the auxiliary estimating function using the observed data $y_{obs}$ and the estimated mapping:
\begin{align}
\theta_{obs,n} &= \arg \max_{\theta}Q(y_{obs};\phi_n(\theta)). \label{eq:SQML}
\end{align}
This is referred to as the simulated quasi-maximum likelihood (SQML) estimator by \citet{Smith1993}, who uses the log-likelihood of the auxiliary model as the estimating function.   \citet{Gourieroux1993} show that the estimator in \eqref{eq:II} is asymptotically more efficient than the one in \eqref{eq:SQML}, provided that an optimal $W$ is chosen.

An estimator that is quite different to the previous two, suggested by \citet{Gallant1996}, involves using the derivative of the estimating function of the auxiliary model.  This is defined for some arbitrary dataset $y$ as
\begin{align*}
S_A(y,\phi) &= \left(\frac{\partial Q(y;\phi)}{\partial \phi_{1}}, \cdots ,\frac{\partial Q(y;\phi)}{\partial \phi_{p_{\phi}}}  \right)^\top,
\end{align*}
where $p_{\phi} = \mathrm{dim}(\phi)$ and $\phi_{i}$ is the $i$th component of the parameter vector $\phi$. The estimator of \citet{Gallant1996} is given by
\begin{align}
\theta_{obs,n} &= \arg \min_{\theta}\left\{\left(\frac{1}{n}\sum_{i=1}^nS_A(y_{i},\phi_{obs})\right)^\top \Sigma \left(\frac{1}{n}\sum_{i=1}^nS_A(y_{i},\phi_{obs})\right)\right\}. \label{eq:EMM}
\end{align} 
The quantity $S_A(y_{obs},\phi_{obs})$ does not appear in \eqref{eq:EMM} as it can be assumed to be 0 by definition.  This approach, referred to by \citet{Gallant1996} as the efficient method of moments (EMM) when the estimating function $Q$ is the auxiliary log-likelihood, can be very computationally convenient if there is an analytic expression for $S_A$, as the method only requires fitting the auxiliary model to data once to determine $\phi_{obs}$ before the II optimisation procedure in equation \eqref{eq:EMM}.  Alternatively, it would be possible to estimate the necessary derivatives, which still could be faster than continually fitting the auxiliary model to simulated data.  The EMM approach is also dependent upon a weighting matrix, $\Sigma$.  One possible choice for $\Sigma$ is $J(\phi_{obs})^{-1}$, since $J(\phi_{obs})$ can be used to estimate the variance of the score. 

\citet{Gourieroux1993} show that the estimators in \eqref{eq:II} and \eqref{eq:EMM} are asymptotically equivalent for certain choices of the weighting matrices.  However, their finite sample performance may differ and thus the optimal choice of estimator may be problem dependent.  For example, \citet{Monfardini1998} compares estimation of a stochastic volatility model using II techniques based on auxiliary autoregressive (AR) and autoregressive moving average (ARMA) models.  Estimation of the AR model is computationally trivial so \citet{Monfardini1998} use the estimator in \eqref{eq:II} whereas the ARMA model is harder to estimate but has an analytic expression for the score thus \eqref{eq:EMM} is used.  A simulation study showed smaller bias for the AR auxiliary model.

As is evident from above, a common estimating function is the auxiliary log-likelihood
\begin{align*}
Q(y_{1:n};\phi) &= \sum_{i=1}^n \log p_A(y_i|\phi).
\end{align*} 
However the user is free to choose the estimating function.  \citet{Heggland2004} demonstrate how the estimating function can involve simple summary statistics of the data, for example the sample moments.  The simulated method of moments \citep{McFadden1989} involves finding a parameter value that generates simulated sample moments closest to the pre-specified observed sample moments.  Thus the simulated method of moments is a special case of II.  \citet{Heggland2004} suggest that the auxiliary statistic should be chosen such that it is sensitive to changes in $\theta$ but is robust to different independent simulated datasets generated based on a fixed $\theta$.  \citet{Heggland2004} also show that when the auxiliary statistic is sufficient for $\theta$ and has the same dimension as $\theta$, the II estimator has the same asymptotic efficiency as the MLE except for a multiplicative factor of $1+1/n$, which reduces to 1 as $n \rightarrow \infty$.  \citet{Jiang2004} mention that II estimators could be improved further via a one step Newton-Raphson correction (e.g. Le Cam 1956), but would require some computations involving the complex generative model. 

The Bayesian indirect inference procedures summarised below are essentially inspired by their classical counterparts above.  Since a number of methods are surveyed here, all with different acronyms, Table \ref{tab:acronyms} defines the acronyms again for convenience together with a description of the methods and key relevant literature.

\begin{table}
	\begin{small}
	\begin{tabular}{c|p{2cm}|p{3.5cm}|p{3.5cm}}
		Acronym & Expansion & Description & Key References \\
		\hline
		ABC II & ABC indirect inference & ABC that uses an auxiliary model to form a summary statistic & \citet{Gleim,DrovandiBII2014} \\
		\hline
		ABC IP & ABC indirect parameter & ABC that uses the parameter estimate of an auxiliary model as a summary statistic & \citet{DrovandiEtAl2011,DrovandiBII2014} \\
		\hline
		ABC IL & ABC indirect likelihood & ABC that uses the likelihood of an auxiliary model to form a discrepancy function & \citet{Gleim,DrovandiBII2014} \\
		\hline
		ABC IS & ABC indirect score & ABC that uses the score of an auxiliary model to form a summary statistic & \citet{Gleim,Martin2014,DrovandiBII2014} \\
		\hline
		BIL & Bayesian indirect likelihood & A general approach that replaces an intractable likelihood with a tractable likelihood within a Bayesian algorithm & \citet{DrovandiBII2014} \\
		\hline
		pdBIL & parametric BIL on the full data level & BIL method that uses the likelihood of a parametric auxiliary model on the full data level to replace the intractable likelihood & \citet{Reeves2005,Gallant2009,DrovandiBII2014} \\
		\hline
		psBIL & parametric BIL on the summary statistic level & BIL method that uses the likelihood of a parametric auxiliary model on the summary statistic level to replace the intractable likelihood of the summary statistic & \citet{DrovandiBII2014} \\
		\hline
		ABC-cp & ABC composite parameter & ABC that uses the parameter of a composite likelihood to form a summary statistic & This chapter but see \citet{Ruli2013} for a composite score approach \\
		\hline
		ABC-ec & ABC extremal coefficients & ABC approach of \citet{Erhardt2012} for spatial extremes models & \citet{Erhardt2012} \\
	\end{tabular}
	\caption{A list of acronyms together with their expansions for the Bayesian likelihood-free methods surveyed in this chapter.  A description of each method is shown together with some key references.}
	\label{tab:acronyms}
	\end{small}
\end{table}

\section{ABC II Methods} \label{sec:ABCII}

The first of the ABC II methods to appear in the literature uses the parameter of the auxiliary model as a summary statistic.  For each dataset that is simulated from the model, $y \sim p(\cdot|\theta)$, the auxiliary model is fitted to this data to produce the simulated summary statistic, $s = \phi_y$.  \citet{DrovandiEtAl2011} propose to compare this simulated summary statistic to the observed summary statistic, $s_{obs} = \phi_{obs}$, which is obtained by fitting the auxiliary model to the observed data prior to the ABC analysis, using the following discrepancy function
\begin{align}
||s-s_{obs}|| &= \sqrt{(\phi_y - \phi_{obs})^\top J(\phi_{obs}) (\phi_y - \phi_{obs})}, \label{eq:ABCIP}
\end{align}  
where the observed information matrix of the auxiliary model evaluated at the observed summary statistic, $J(\phi_{obs})$, is utilised to provide a natural weighting of the summary statistics that also takes into account any correlations between the components of the auxiliary parameter estimate.  This method is referred to by \citet{DrovandiBII2014} as ABC IP.  Instead, if we base the discrepancy on the auxiliary likelihood \citep{Gleim}, we obtain the ABC IL method
\begin{align*}
||s-s_{obs}|| &= \log p_A(y_{obs}|\phi_{obs}) - \log p_A(y_{obs}|\phi_y).
\end{align*} 
Under some standard regularity conditions it is interesting to note that the ABC IP discrepancy function appears in the second order term of the Taylor series expansion of the ABC IL discrepancy function.  This might suggest that the ABC IL discrepancy function is more efficient than the discrepancy function of ABC IP generally but this requires further investigation.  A common aspect of the ABC IP and ABC IL approaches is that they both use the auxiliary parameter estimate as a summary statistic.  As such, these methods involve fitting the auxiliary model to every dataset simulated from the generative model during an ABC algorithm.  In one sense, it is desirable to extract as much information out of the auxiliary model as possible.  From this point of view, an attractive estimation procedure is maximum likelihood
\begin{align*}
\phi_{obs} = \arg \max_{\phi \in \Phi}p_A(y_{obs}|\phi),
\end{align*}
which tends to be more efficient than other simpler estimation approaches such as the method of moments.  However, in most real applications there is not an analytic expression for the auxiliary MLE, and one must then resort to numerical optimisation algorithms (e.g.\ \citet{DrovandiEtAl2011} apply the Nelder-Mead derivative free optimiser).  Having to determine a numerical MLE at every iteration of the ABC algorithm not only slows down the method, but also potentially introduces further issues if the numerical optimiser is prone to getting stuck at local modes of the auxiliary likelihood surface.  A pragmatic approach may be to initialise the numerical optimiser at the observed auxiliary parameter estimate, $\phi_{obs}$.  

In summary, the above review reveals that the optimal choice of auxiliary estimator may be a trade-off between the computational cost of obtaining and the statistical efficiency of the chosen auxiliary estimator.  One approach to expand on this literature might be to start with a computationally simple but consistent estimator (e.g.\ the method of moments) and apply one iteration of a Newton-Raphson method to produce an asymptotically efficient estimator \citep{LeCam1956} in a timely manner.  It is important to note that the ABC IP and ABC IL methods are essentially ABC versions of the classical II approaches in \citet{Gourieroux1993} (who compare observed and simulated auxiliary parameters) and \citet{Smith1993} (who maximise the auxiliary log-likelihood), respectively.

A rather different approach to ABC IP and ABC IL considered by \citet{Gleim} uses the score of the auxiliary model as the summary statistic, resulting in the ABC IS procedure.   A major advantage of the ABC IS approach is that we always evaluate the score at the observed auxiliary estimate, $\phi_{obs}$.  Any simulated data, $y$, obtained during the ABC algorithm can be substituted directly into the auxiliary score to determine the simulated summary statistic, without needing to fit the auxiliary model to the simulated data.  In cases where there is an analytic expression for the score, the summary statistic can be very fast to compute (similar to more traditional summary statistics used in ABC) and this leads to substantial computational savings over ABC IP and ABC IL.  However, in many applications, the derivatives of the auxiliary likelihood are not available analytically.  In such situations it is necessary to estimate the derivatives numerically (see, for example, \citet{Martin2014} and Section \ref{subsec:Infectious}) using a finite difference strategy, for example.  This may contribute another small layer of approximation and add to the computational cost, although the number of likelihood evaluations required to estimate the score is likely to be less than that required to determine the auxiliary MLE.  

When the MLE is chosen as the auxiliary estimator then the observed score (or summary statistic here, $s_{obs}$) can be assumed to be numerically $0$.  Therefore a natural discrepancy function in the context of ABC IS is given by
\begin{align}
||s-s_{obs}|| &= \sqrt{S_A(y,\phi_{obs})^\top J(\phi_{obs})^{-1}S_A(y,\phi_{obs})}, \label{eq:score_abc}
\end{align}     
where $s = S_A(y,\phi_{obs})$. We can again utilise the observed auxiliary information matrix to obtain a natural weighting of the summary statistics.  The ABC IS approach is effectively an ABC version of the EMM method in \citet{Gallant1996}.

The assumptions required for each ABC II approach to behave in a satisfactory manner are provided in \citet{DrovandiBII2014}.  In summary, the ABC IP approach requires a unique auxiliary parameter estimator so that each simulated dataset results in a unique value of the ABC discrepancy.  ABC IL requires a unique maximum likelihood value and ABC IS requires a unique score (and some other mild conditions) for each simulated dataset generated during the ABC algorithm.  \citet{DrovandiBII2014} consider an example where the auxiliary model is a mixture model, which does not possess a unique estimator due to the well-known label switching issue with mixtures.  The ABC IL and ABC IS approaches were more suited to handling the label switching problem.

\citet{Martin2014} contain an important result that shows that the auxiliary score carries the same information as the auxiliary parameter estimate, thus the ABC II approaches will have the same target distribution in the limit as $h \rightarrow 0$.  \citet{Martin2014} demonstrate that the discrepancy function involving the auxiliary score can be written as a discrepancy function involving the auxiliary parameter estimate, thus ABC will produce the same draws regardless of the choice of summary statistic in the limit as $h \rightarrow 0$.  Whilst \citet{DrovandiBII2014} demonstrate empirically that there are differences amongst the ABC II results for $h>0$, the result of \citet{Martin2014} does provide more motivation for a score approach, which will often be more computationally efficient.

In choosing an auxiliary model in the context of ABC II, it would seem desirable if the auxiliary model gave a good fit to the observed data so that one is reasonably confident that the quantities derived from the auxiliary model capture most of the information in the observed data.  In particular, the auxiliary MLE is asymptotically sufficient for the auxiliary model.  Thus, assuming some regularity conditions on the auxiliary model, if the generative model is a special case of the auxiliary model then the statistic derived from the auxiliary model will be asymptotically sufficient also for the generative model \citep{Gleim}.  An advantage of the ABC II approach is that the utility of the summary statistic may be assessed prior to the ABC analysis by performing some standard statistical techniques such as goodness-of-fit and/or residual analyses.  For example, \citet{DrovandiBII2014} consider a chi-square goodness-of-fit test to indicate insufficient evidence against the auxiliary model providing a good description of the data in an application involving a stochastic model of macroparasite population evolution.  This is in contrast to more traditional choices of summary statistics \citep{Blum2013}, where it is often necessary to perform an expensive simulation study to select an appropriate summary.  The well-known curse of dimensionality issue associated with the choice of summary statistic in ABC \citep{Blum2009} can be addressed to an extent in the ABC II approach by choosing a parsimonious auxiliary model which might be achieved by comparing competing auxiliary models through some model selection criterion (for example, \citet{DrovandiEtAl2011} use the Akaike Information Criterion).

Alternatively it might be more convenient to select the auxiliary model as a simplified version of the generative model so that the auxiliary parameter has the same interpretation as the generative parameter (parameter of the generative model).  For example, Section \ref{subsec:Infectious} considers performing inference for a Markov process using the corresponding linear noise approximation as the auxiliary model.  This approach has the advantage that the summary statistic will have the same dimension as the parameter of interest (which can be desirable, see e.g.\ \citet{Fearnhead2012}, and indeed necessary for some methods \citep{Nott2014,Li2015}) and that there will be a strong connection between the generative and auxiliary parameters.  The obvious drawback is, assuming the generative model is correct, the simplified version of the model in general will not provide a good fit to the observed data, and ultimately any asymptotic sufficiency for the auxiliary model is lost for the generative model.   

From a classical point of view, it is well known that basing inferences on the wrong model may result in biased parameter estimates.  From a Bayesian perspective, the mode and concentration of the posterior distribution may not be estimated correctly when employing an approximate model for inference purposes.  As mentioned in Section \ref{sec:II}, II can be viewed as a method that provides some correction for assuming the wrong/simplified version of the model \citep{Jiang2004}.  It does this by finding the parameter value of the generative model that leads to simulated data where the parameter estimate based on the simplified model applied to the simulated data is closest to that of the observed data.  It may also be that applying ABC II in a similar way may lead to posterior modes that are closer to the true posterior mode in general compared to when inferences are solely based on the misspecified model.  Furthermore, ABC has a tendency to provide a less concentrated posterior distribution relative to the true posterior, which depends on how much information has been lost in the data reduction.  Thus, using auxiliary parameter estimates of a simplified model as summary statistics in ABC will not lead to over concentrated posteriors, as may be the case if the simplified model was used directly in the Bayesian analysis.   Using a summary statistic derived from such an auxiliary model in ABC is yet to be thoroughly explored in the literature (although see \citet{Martin2014} for an example). 

Under certain regularity conditions on the auxiliary model that lead to II producing a consistent estimator (e.g.\ \citet{Gourieroux1993}), ABC II will produce Bayesian consistency in the limit as $h \rightarrow 0$ \citep{Martin2014}.  Under the regularity conditions, in equation \eqref{eq:ABCIP}, $\phi_{obs} \rightarrow \phi(\theta_0)$ (where $\theta_0$ is the true value of the parameter) and $\phi_y \rightarrow \phi(\theta)$ where $y \sim p(y|\theta)$ as the sample size goes to infinity.  Thus in the limit as $h \rightarrow 0$, ABC will only keep $\theta = \theta_0$.

\citet{Martin2014} also take advantage of the strong one-to-one correspondence between the auxiliary and generative parameters in the situation where the auxiliary model is a simplified version of the generative model.  Here, \citet{Martin2014} suggest the use of the marginal score for a single auxiliary parameter, which is the score with all other components of the parameter integrated out of the auxiliary likelihood, as a summary statistic to estimate the univariate posterior distribution of the corresponding single generative model parameter.

\section{BIL with a Parametric Auxiliary Model} \label{sec:BIL}

An alternative to ABC II that has been considered in the literature by \citet{Reeves2005} and \citet{Gallant2009} is to use the likelihood of an auxiliary parametric model as a replacement to that of the intractable generative likelihood provided a relationship, $\phi(\theta)$, has been estimated between the generative and auxiliary parameters.  See also \citet{Ryan2014} for an application of this method to Bayesian experimental design in the presence of an intractable likelihood.  This approach is investigated in more theoretical detail in \citet{DrovandiBII2014} and is referred to as pdBIL (where d stands for data).  It is also possible to apply a parametric auxiliary model to the summary statistic likelihood, where some data reduction has been applied.  One then obtains psBIL (where s denotes summary statistic), which is discussed briefly later in this section.  The pdBIL method could be considered as a Bayesian version of the SQML approach in \citet{Smith1993}.  If the so-called mapping or binding function, $\phi(\theta)$, is known, then the approximate posterior of the pdBIL approach is given by
\begin{align*}
\pi_{A}(\theta|y_{obs})\propto p_{A}(y_{obs}|\phi(\theta))\pi(\theta),
\end{align*}  
where $\pi_{A}(\theta|y_{obs})$ denotes the pdBIL approximation to the true posterior and $p_{A}(y_{obs}|\phi(\theta))$ is the likelihood of the auxiliary model evaluated at the parameter $\phi(\theta)$.  

Unfortunately, in practice, the relationship between $\phi$ and $\theta$ will be unknown.  More generally, it is possible to estimate $\phi(\theta)$ via simulation of $n$ independent and identically distributed datasets, $y_{1:n} = (y_1,\ldots,y_n)$, from the generative model based on a proposed parameter, $\theta$.  Then the auxiliary model is fitted to this large dataset to obtain $\phi_n(\theta)$, which could be based on maximum likelihood
\begin{align*}
\phi_n(\theta) = \arg \max_{\phi} \prod_{i=1}^np_A(y_{i}|\phi).
\end{align*}
The target distribution of the resulting method is given by
\begin{align*}
\pi_{A,n}(\theta|y_{obs}) \propto p_{A,n}(y_{obs}|\theta)\pi(\theta),
\end{align*}
where
\begin{align*}
p_{A,n}(y_{obs}|\theta)=\int_{y_{1:n}}p_{A}(y_{obs}|\phi_{n}(\theta))\left\{\prod_{i=1}^{n}p(y_{i}|\theta)\right\}{\rm d}y_{1:n},
\end{align*}  
which can be estimated unbiasedly using a single draw of $n$ iid datasets, $y_{1:n} \sim p(\cdot|\theta)$.  The introduction of the second subscript $n$ in $p_{A,n}(y_{obs}|\theta)$ highlights that an additional layer of approximation is introduced by selecting a finite value of $n$ to estimate the mapping (see \citet{DrovandiBII2014} for more details).  The empirical evidence in \citet{DrovandiBII2014} seems to suggest, in the context of pdBIL, that the approximate posterior becomes less concentrated as the value of $n$ decreases.  Therefore, if the auxiliary model chosen is reasonable (discussed later in this section), then better posterior approximations can be anticipated by taking $n$ as large as possible.  Initially it would seem apparent that the computational cost of the pdBIL approach would grow as $n$ is increased.  However, \citet{DrovandiBII2014} report an increase in acceptance probability of an MCMC algorithm targeting $\pi_{A,n}(\theta|y_{obs})$ as $n$ is increased.  Thus, up to a point, $n$ can be raised without increasing the overall computing time since fewer iterations of MCMC will be required to obtain an equivalent effective sample size compared with using smaller values of $n$.  Values of $n$ above a certain limit where the acceptance probability does not increase may reduce the overall efficiency of the approach.

Like ABC IP and ABC IL, pdBIL requires an optimisation step for every simulated dataset so it can be an expensive algorithm.  For a Potts model application with a single parameter, \citet{Moores2014} propose to run pre-simulations across the prior space for a chosen value of $n$ and fit a non-parametric model in order to smooth out the effect of $n$ and recover an estimate of the mapping, denoted $\hat{\phi}(\theta)$.  A major computational advantage of this approach is that the estimated mapping can be re-used to analyse multiple observed datasets of the same size.  However, devising a useful strategy to extend this idea to higher dimensional problems than that considered by \citet{Moores2014} is an open area of research.   

For the pdBIL method to lead to a quality approximation, the approach relies on a quite strong assumption that the auxiliary likelihood acts as a useful replacement likelihood across the parameter space with non-negligible posterior support and that the auxiliary likelihood reduces further in regions of very low posterior support \citep{DrovandiBII2014}.  In contrast, the ABC II methods require that the summary statistic coming from the auxiliary model is informative and an efficient algorithm is available to ensure a close matching between the observed and simulated summary statistics in order to produce a close approximation to the true posterior distribution.  \citet{DrovandiBII2014} demonstrate that under suitable conditions the pdBIL method will target the true posterior in the limit as $n \rightarrow \infty$ if the generative model is nested within the auxiliary model.  In this ideal scenario, ABC II methods will not be exact as $h \rightarrow 0$ since the quantities drawn from the auxiliary model will not produce a sufficient statistic in general, as the dimension of the statistic will be smaller than the size of the data (however, under suitable regularity conditions, the statistic will be asymptotically sufficient \citep{Gleim}).  Of course, it would seem infeasible to find a tractable auxiliary model that incorporates an intractable model as a special case.  However, this observation does suggest, in the context of pdBIL method, that a flexible auxiliary model may be useful.

We note that pdBIL is not illustrated empirically in this chapter but a number of examples are provided in \citet{Reeves2005,Gallant2009,DrovandiBII2014}.

We note that a parametric auxiliary model can also be applied at a summary statistic level; that is, when some data reduction technique has been performed. As mentioned earlier, the method is referred to as psBIL in \citet{DrovandiBII2014}.  A popular psBIL method in the literature is to assume a multivariate normal auxiliary model.  Here the likelihood of the multivariate normal distribution, with a mean and covariance matrix dependent on $\theta$, is used as a replacement to the intractable summary statistic likelihood.  This technique has been referred to as synthetic likelihood \citep{Wood2010,Price2018}, and is covered in much greater detail in another chapter \citep{Drovandi2018}.

\section{Further Reading} \label{sec:Further}

\citet{Ruli2013} propose to use the score of a composite likelihood approximation of the full likelihood as a summary statistic for ABC (referred to as ABC-cs).  Methods involving the composite likelihood can be applied when the full data likelihood is intractable but the likelihood of certain subsets of the data can be evaluated cheaply.  Section \ref{subsec:spatial} considers an example in spatial extremes where composite likelihood methods are applicable.  The method of \citet{Ruli2013} has a strong connection with the ABC IS method but it does not quite fall under the ABC II framework as the composite likelihood is not associated with any parametric auxiliary model, but rather is used as a proxy to the full data likelihood formed by simplifying the dependency structure of the model.  Of course, an alternative to \citet{Ruli2013} could use the composite likelihood estimate as a summary statistic and obtain approaches similar to ABC IP and ABC IL (Section \ref{subsec:spatial} considers a composite likelihood variant on ABC IP).  However, as with ABC IS, the approach of \citet{Ruli2013} can be relatively fast if the composite score is easier to obtain than the composite parameter estimator.

\citet{PauliEtAl2011} and \citet{CooleyEtAl2011} consider a Bayesian analysis where they use the composite likelihood directly as a replacement to the true likelihood.  A naive application of this can lead to posterior approximations that are incorrectly concentrated as each data point may appear in more than one composite likelihood component depending on how the data subsets are constructed.  However,  \citet{PauliEtAl2011} and \citet{CooleyEtAl2011} suggest so-called calibration approaches in an attempt to correct this.  The method of \citet{Ruli2013} essentially by-passes this issue by using the composite likelihood to form a summary statistic for ABC.  Therefore, \citet{Ruli2013} rely on this summary statistic being approximately sufficient in order to achieve a good approximation to the true posterior distribution.  The approach of \citet{Ruli2013} generally falls under the Bayesian indirect inference framework as it involves simulation from the true model in an attempt to correct an estimator  based solely on the composite likelihood.  The dimension of the summary statistic will coincide with that of the generative model parameter and there will likely be a strong correspondence between the two.    

\citet{Forneron2016} develop an approach called the reverse sampler (RS) that produces approximate Bayesian inferences that has a strong connection with II.  The approach involves solving many II problems with a different random seed each time, and upon re-weighting the resulting solutions, an independent sample is generated from an approximate posterior.  The summary statistic may come from an auxiliary model or could be any summarisation of the full data.  The approach is provided in Algorithm \ref{alg:RS}.  It is important to note that each II optimisation uses $n=1$, as in ABC.  For this approach we denote the simulated summary statistic as $s(\theta,\xi)$ where $\xi$ are a set of random numbers generated through the simulation, $\xi \sim p(\xi)$.  For each II optimisation procedure, $\xi$ is held fixed.  This is equivalent to using the same random seed during each II optimisation.

Denote the sample obtained from solving the $i$th optimisation problem as $\theta^{(i)}$.  After $\theta^{(i)}$ is generated it must be weighted.  One aspect of the weighting is the prior density, $\pi(\theta^{(i)})$.  It also involves a Jacobian term and a volume term if the number of summary statistics exceeds the dimension of the parameter.  Denote $s_\theta(\theta,\xi)$ as the Jacobian
\begin{align*}
s_\theta(\theta,\xi) &= \frac{\partial s(\theta,\xi)}{\partial \theta},
\end{align*}
which is a $q\times p$ matrix where $q$ is the dimension of the summary statistic and $p$ is the dimension of the parameter.  That is, the $(j,k)$ element of this matrix is given by $\frac{\partial s_j(\theta,\xi)}{\partial \theta_k}$ where $s_j(\theta,\xi)$ is the function for the $j$th summary statistic.  Then the weight for sample $\theta^{(i)}$ is given by
\begin{align*}
w^{(i)} \propto \pi(\theta^{(i)})\mbox{vol}(s_\theta(\theta^{(i)},\xi^{(i)}))^{-1},  
\end{align*} 
where
\begin{align*}
\mbox{vol}(s_\theta(\theta^{(i)},\xi^{(i)})) &= \sqrt{\mbox{det}\left( s_\theta(\theta^{(i)},\xi^{(i)})^\top s_\theta(\theta^{(i)},\xi^{(i)}) \right)}.
\end{align*}
Upon normalisation of the weights, a weighted sample is obtained from an approximate posterior.  \citet{Forneron2016} also include a kernel function to the weights, $K_h(||s_{obs}-s_\theta(\theta^{(i)},\xi^{(i)})||)$ to give higher weight to II optimisation samples that get closer to the observed summary statistic.  Note that if $s_\theta(\theta^{(i)},\xi^{(i)})$ is unavailable analytically it can be estimated via numerical differentiation, for example, finite differencing.

\begin{algorithm}
	\caption{The reverse sampler of \citet{Forneron2016}.}
	\label{alg:RS}
	\begin{algorithmic}[1]
		\FOR{$i=1$ \TO $T$ where $T$ is the number of samples}
		\STATE Solve the II optimisation problem $\theta^{(i)} = \arg \max_{\theta}\{(s(\theta,\xi^{(i)})-s_{obs})^\top W (s(\theta,\xi^{(i)})-s_{obs})\}$ where $\xi^{(i)} \sim p(\xi)$.  Set $\rho^{(i)} = (s(\theta^{(i)},\xi)-s_{obs})^\top W (s(\theta^{(i)},\xi)-s_{obs})$
		\STATE Set the weight for sample $\theta^{(i)}$ as $w^{(i)} \propto \pi(\theta^{(i)})\mbox{vol}(s_\theta(\theta^{(i)},\xi^{(i)}))$  
		\ENDFOR
	\end{algorithmic}
\end{algorithm} 

Here we provide an example to obtain some insight into the approximation behaviour of the RS.  Consider a dataset $y_1,\ldots,y_N \sim N(\mu,1)$ of length $N$.  A sufficient statistic is $s = \bar{y} \sim N(\mu,1/\sqrt{n})$. We may think of the simulated data as a transformation of the parameter and some noise variables, $y_i = \mu + \xi_i$ where $\xi_i \sim N(0,1)$ and $\xi = (\xi_1,\ldots,\xi_N)$.  To obtain the summary statistic we have $s(\mu,\xi) = \bar{y} = \mu + \bar{\xi}$ where $\bar{\xi}$ is the sample mean of the noise variables.  The Jacobian of this transformation is 1.  For fixed $\bar{\xi}^{(i)}$ we obtain $\mu^{(i)} = \bar{y}_{obs} - \bar{\xi}^{(i)}$, in which case $s_{obs} - s(\mu^{(i)},\xi^{(i)}) = 0$.   Effectively the RS algorithm samples from $\mu \sim N(\bar{y}_{obs},1/\sqrt{n})$ which is the posterior distribution of $\mu$ if the prior is uniform and improper over the real line.  If the prior was selected differently then the $\mu^{(i)}$ samples need to be weighted proportional to the prior density $w^{(i)} \propto \pi(\mu^{(i)})$.  Assume for the rest of this example that $\pi(\mu) \propto 1$ for $-\infty < \mu < \infty$.

Now consider the two-dimensional summary $(\bar{y},m)$ where $m$ is the sample median.  This remains a sufficient statistic, so that ABC targets the true posterior in the limit as $h \rightarrow 0$.  Here we have $\bar{y} = \mu + \bar{\xi}$ and $m = \mu + m_\xi$ where $\bar{\xi}$ is the sample mean of the noise variables and $m_\xi$ is the median of the noise variables.  We set the discrepancy function as the following
\begin{align*}
||s_{obs} - s(\mu,\xi)|| = \left\{\bar{y}_{obs} - (\mu + \bar{\xi})\right\}^2 + \left\{m_{obs} - (\mu + m_\xi) \right\}^2.
\end{align*}

\begin{figure}[!htp]
	\centering
	\subfigure[]{\includegraphics[height=0.3\textheight,width=0.45\textwidth]{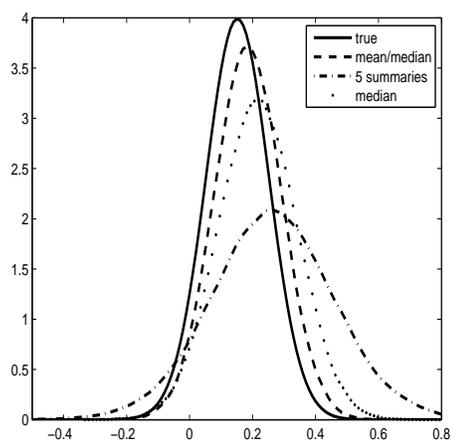}\label{subfig:rs_summaries}}
	\subfigure[]{\includegraphics[height=0.3\textheight,width=0.45\textwidth]{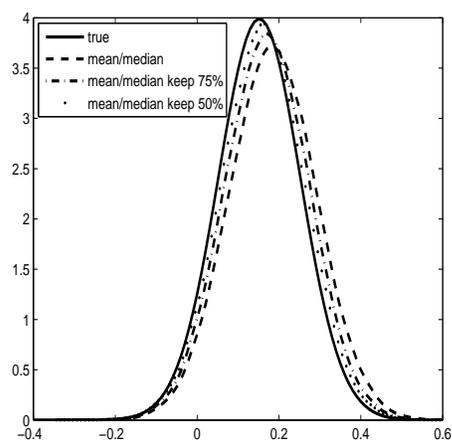}\label{subfig:rs_summaries_keep2}}
	\subfigure[]{\includegraphics[height=0.3\textheight,width=0.45\textwidth]{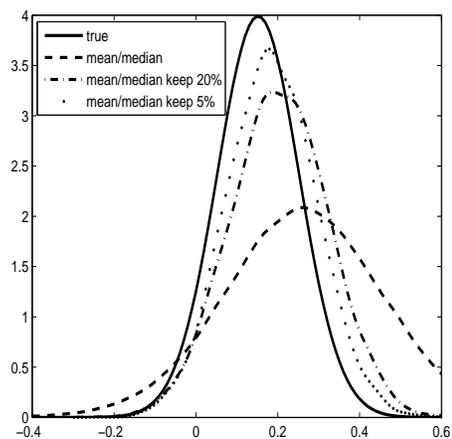}\label{subfig:rs_summaries_keep5}}
	\subfigure[]{\includegraphics[height=0.3\textheight,width=0.45\textwidth]{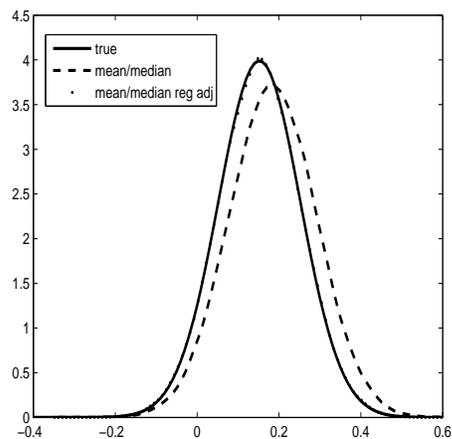}\label{subfig:rs_summaries_regadj2}}
	\subfigure[]{\includegraphics[height=0.3\textheight,width=0.45\textwidth]{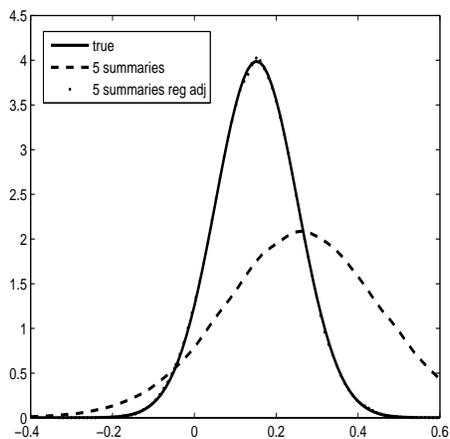}\label{subfig:rs_summaries_regadj5}}
	\caption{RS example for the normal location model with different choices of the summary statistics and using rejection and regression adjustment.}
	\label{fig:toy}
\end{figure}

Minimising this with respect to $\mu$ for some $\xi^{(i)}$ we obtain $\mu^{(i)} = (\bar{y}_{obs} - \bar{\xi}^{(i)} + m_{obs} - m_\xi^{(i)})/2$.  Thus in this case RS does not draw directly from the posterior (despite the fact that the statistic is sufficient).  This is confirmed in Figure \ref{subfig:rs_summaries} where there is departure from the true posterior (although it is better than an RS sampler that just uses the median as a summary).  Here we investigate the impact on the approximate posterior as we discard some of the $\mu$ RS samples with the highest discrepancy.  Figure \ref{subfig:rs_summaries_keep2} demonstrates improved accuracy by discarding only 50\% of the draws.  This is equivalent to using a uniform kernel with an appropriate value of $h$ in the RS weighting function of \citet{Forneron2016}.

Also shown in Figure \ref{subfig:rs_summaries} is the posterior when using 5 summary statistics (mean, median, min, max, midrange).  The partial derivatives of the summaries (for fixed noise variables) with respect to the parameter are all equal to 1 so that the Jacobian term for all samples is the same so that the weights are the same for each sample.  Again using the squared Euclidean distance it is easy to show what the expression is for $\mu$ drawn during the RS sampler.  Here there is quite a large departure from the true posterior, demonstrating that the RS method does suffer from a curse of dimensionality in the summary statistic, as with ABC.  Figure \ref{subfig:rs_summaries_keep5} shows that the approximation can be improved by discarding samples with the highest discrepancy values.   However, a large proportion (around 95\% say) need to be discarded to get close to the true posterior.  The results demonstrate that care must be taken when choosing summary statistics for RS, as in ABC.

Figures \ref{subfig:rs_summaries_regadj2} and \ref{subfig:rs_summaries_regadj5} demonstrate that using a local regression adjustment of \citet{Beaumont2002} (as is commonly done in ABC) almost fully corrects the RS approximations.  It appears the regression adjustment can be quite useful for RS, as it is for ABC.

\section{Examples} \label{sec:Examples}

In this section we consider two case studies involving real data to demonstrate how ideas from indirect inference can enhance ABC inferences.  Note throughout all ABC analyses we use the uniform kernel weighting function, $K_h(||s-s_{obs}||) = I(||s-s_{obs}||\leq h)$.

\subsection{Infectious Disease Example} \label{subsec:Infectious}

Consider a stochastic susceptible-infected (SI) model where the number of susceptibles and infecteds at time $t$ are denoted by $S(t)$ and $I(t)$ respectively.  In an infinitesimal time $\Delta_t$ a transmission can occur with approximate probability $\beta S(t) I(t) \Delta_t$, which increments the number of infectives and reduces the number of susceptibles by one.  A removal can occur in that time with approximate probability $\gamma I(t)\Delta_t$.  Removed individuals play no further part in the process.  This model has been applied to a smallpox epidemic dataset in \citet{ONeill1999,Fearnhead2004}.  The dataset consists of the removal times in days (see \citet[pg.\ 111]{Becker1989}).  The village consists of 120 people, one of whom becomes infected and introduces it into the community.  We set the first removal at time $t=0$, where for simplicity we assume that there is exactly one infected, i.e.\ $I(0) = 1$ and $S(0) = 118$.  We adopt the same interpretation of the dataset that \citet{Golightly2014} use where the observations refer to daily recordings of $S(t) + I(t)$, which changes only on days where a removal takes place.  Thus we do not assume that the epidemic has necessarily ceased with the last removal recorded in the dataset.

Samples from the true posterior distribution can be obtained using the particle MCMC approach specified in \citet{DrovandiCountTS2014} that makes use of the alive particle filter (see \citet{Jasra2013} for additional information on the alive particle filter).  However, in general, it is very difficult to devise exact and computationally feasible strategies for these Markov process models, especially when the model contains several populations.  Therefore here we consider approximate Bayesian inference approaches that use a direct approximation to the true stochastic process.  A popular and tractable approximation to continuous time Markov chains is the linear noise approximation (LNA).  \citet{Fearnhead2014} base their posterior inferences directly on the LNA and thus require that the LNA be a very good approximation to the Markov process, which may not always be the case.

A different approach is to use the auxiliary LNA model to form summary statistics for ABC. In this situation there are as many summary statistics as generative model parameters and the auxiliary and generative model parameters have the same interpretation.  We denote the LNA model parameters as $\beta^{A}$ and $\gamma^A$, where superscript $A$ denotes the auxiliary model.  The auxiliary model likelihood is denoted as $p_A(y|\phi)$ where $\phi = (\beta^{A},\gamma^A)$ is the auxiliary parameter.  The auxiliary parameter estimate based on the observed data is given by
\begin{align*}
\phi_{obs} &= \arg \max_{\phi}p_A(y_{obs}|\phi).
\end{align*}

To reduce the computational burden, we consider an ABC IS approach.  That is, we consider the score of the LNA model and always use $\theta^A_{obs}$ in the score function for the summary statistic:
\begin{align*}
s = S_A(y,\phi_{obs}) &= \left(\frac{\partial \log p_A(y|\phi)}{\partial \beta^{A}}|_{\phi=\phi_{obs}}, \frac{\partial \log p_A(y|\phi)}{\partial \gamma^A}|_{\phi=\phi_{obs}}  \right)^\top,
\end{align*}  
where $y$ is a simulated dataset.  We assume that $\phi_{obs}$ has been obtained accurately enough so that we can assume $s_{obs} = S_A(y_{obs},\phi_{obs}) = (0,0)^\top$.  We set the discrepancy function $||\cdot||$ as the $L_2$ norm of $s$.  However, there is no analytic expression for the score.  Therefore we estimate the score numerically, which requires several evaluations of the LNA likelihood.  Thus calculating the summary statistic based on the score is mildly computationally intensive, and given that ABC tends to suffer from poor acceptance rates, we propose a method here to accelerate the algorithm without altering the ABC target distribution.

To improve the computational efficiency of the ABC approach we propose an implementation of the lazy ABC method of \citet{Prangle2014}.  Our approach uses a second summary statistic, which we call the lazy summary statistic (denoted by $s_{obs,\mathrm{lazy}}$ and $s_{\mathrm{lazy}}$ for the observed and simulated data, respectively), for which computation is trivial once data have been simulated from the generative model but which is less informative than the originally proposed summary statistic.  The discrepancy function for the lazy summary statistic is given by
\begin{align*}
\rho_{\mathrm{lazy}} = ||s_{\mathrm{lazy}}-s_{obs,\mathrm{lazy}}||.
\end{align*}

Firstly, a decision on whether or not to compute the actual summary statistic is made on the basis of the distance between the observed and simulated lazy summary statistics.  If $\rho_{\mathrm{lazy}}$ falls below a threshold, $h_{\mathrm{lazy}}$, then the proposal may have produced simulated data reasonably close to the observed data and it is then worthwhile to compute the more informative summary statistic.  However, if the lazy distance is too high, then the proposed parameter may be rejected and thus calculation of the expensive summary is not required.  In order to obtain an algorithm that preserves the original ABC target, it is necessary to include a continuation probability, $\alpha$, which may require some tuning. Therefore a proposal that performs poorly in terms of the lazy summary statistic will still make it through to the second stage with probability $\alpha$.  We implement this approach within an MCMC ABC method (\citet{Prangle2014} consider ABC importance sampling), which must include $\alpha$ in the acceptance probability to ensure a theoretically correct method.  When $\rho_{\mathrm{lazy}} > h_{\mathrm{lazy}}$ we effectively estimate the ABC likelihood with a random variable that can be two possible values:
\begin{align*}
\begin{array}{cc}
0 & \mbox{with probability } 1-\alpha \\
\frac{I(||s-s_{obs}||\leq h)}{\alpha} & \mbox{with probability } \alpha
\end{array} .
\end{align*}   
The expected value of this random variable is $I(||s-s_{obs}||\leq h)$, the desired ABC likelihood.  Thus our implementation of the lazy ABC approach is an instance of the pseudo-marginal method of \citet{Andrieu2009}.  The drawback of this approach is that it inflates the variance of the ABC likelihood.  If $\rho_{\mathrm{lazy}} > h_{\mathrm{lazy}}$ and $\rho \leq h$ then the ABC likelihood gets inflated, meaning that the MCMC algorithm may become stuck there.

We also incorporate the early rejection strategy of \citet{Picchini2014}. The approach of \citet{Picchini2014} involves a simple re-ordering of steps in MCMC ABC where it is possible to reject a proposed parameter prior to simulating data.  This approach does not alter the target distribution when a uniform weighting function is applied, as we do here.

For this application, the lazy summary statistic is a scalar so we choose the absolute value to compute the distance between the observed and simulated lazy summary statistic in line 7 of Algorithm \ref{alg:MCMC}.  For the actual summary statistic we use the discrepancy function in equation \eqref{eq:score_abc} at line 18 of Algorithm \ref{alg:MCMC}, with $J$ set as the identity matrix for simplicity.

\begin{algorithm}
	\caption{MCMC ABC algorithm using a lazy summary statistic.}
	\label{alg:MCMC}
	\begin{algorithmic}[1]
		\STATE Set $C=1$
		\STATE Set $\theta^{(0)}$ 
		\FOR{$i=1$ \TO $T$}
		\STATE Draw $\theta^* \sim q(\cdot|\theta^{(i-1)})$
		\STATE Compute $r = \min\left(1,\frac{\pi(\theta^*)}{C\pi(\theta^{(i-1)})}\right)$
		\IF {$U(0,1) < r$}
		\STATE Simulate $y \sim p(\cdot|\theta^*)$
		\STATE Compute lazy ABC discrepancy $\rho_{\mathrm{lazy}} = ||s_{\mathrm{lazy}}-s_{obs,\mathrm{lazy}}||$
		\IF {$\rho_{\mathrm{lazy}}$ $> h_{\mathrm{lazy}}$}
		\IF {$U(0,1)$ $< \alpha$}
		\STATE Continue and set $C_{\mathrm{prop}} = 1/\alpha$
		\ELSE
		\STATE Reject early: set $\theta^{(i)} = \theta^{(i-1)}$ and go to the next iteration of the MCMC algorithm
		\ENDIF
		\ELSE
		\STATE Set $C_{\mathrm{prop}} = 1$
		\ENDIF
		\STATE Compute ABC discrepancy $\rho = ||s-s_{obs}||$
		\IF {$\rho \leq h$}
		\STATE $\theta^{(i)} = \theta^*$ and $C = C_{\mathrm{prop}}$
		\ELSE
		\STATE $\theta^{(i)} = \theta^{(i-1)}$
		\ENDIF
		\ELSE
		\STATE $\theta^{(i)} = \theta^{(i-1)}$
		\ENDIF
		\ENDFOR
	\end{algorithmic}
\end{algorithm}

\begin{figure}
	\centering
	\subfigure[$\log \beta$]{\includegraphics[height=0.3\textheight,width=0.45\textwidth]{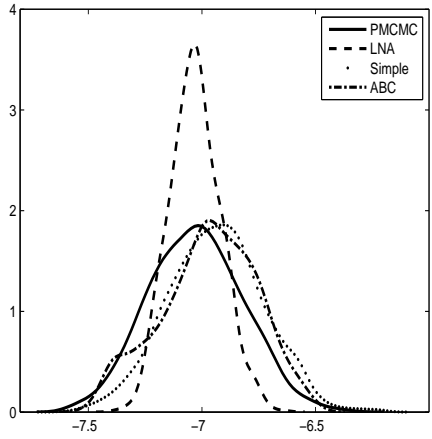}\label{figsub:post_epidemic_beta}}
	\subfigure[$\log\gamma$]{\includegraphics[height=0.3\textheight,width=0.45\textwidth]{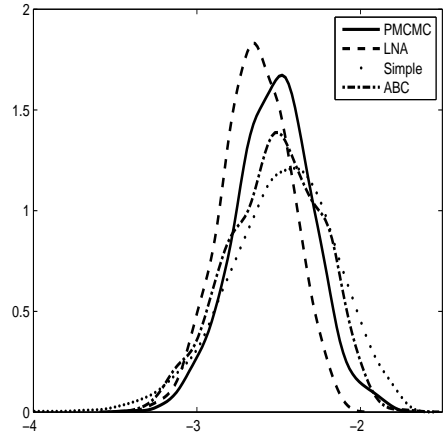}\label{figsub:post_epidemic_gamma}}
	\subfigure[pilot]{\includegraphics[height=0.3\textheight,width=0.45\textwidth]{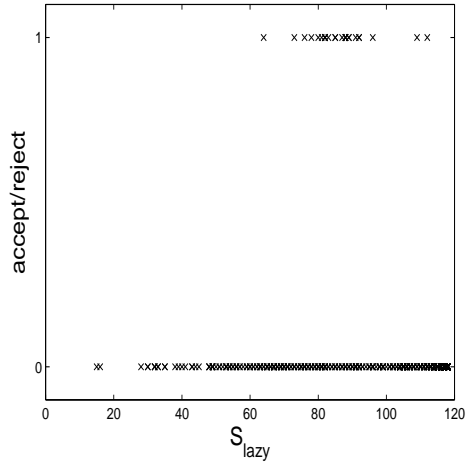}\label{figsub:summ_stat_lazy_acc}}
	\caption{Results for the epidemic example.  Shown are the posterior distributions of (a) $\log \beta$ and (b) $\log \gamma$ based on the particle MCMC approach (solid), LNA approximation (dash), ABC with the final observation as the summary statistic (dot) and ABC with the LNA parameter as a summary statistic (dot-dash).  Subfigure (c) shows the results of the pilot run where the x-axis is the value of the lazy summary statistic whereas the y-axis shows the ABC accept/reject outcome based on the corresponding summary statistic formed by the LNA auxiliary model.}
	\label{fig:post_epidemic}
\end{figure}

The results of using the LNA approximation and the ABC methods are shown in Figure \ref{fig:post_epidemic} (note that the ABC results are based on the lazy ABC implementation).  The results are compared with particle MCMC, which has the true posterior as its limiting distribution.  The LNA results tend to be overprecise (especially for $\beta$) whereas the ABC results tend to be slightly conservative.  Note that we also ran ABC using the final observation as the (simple) summary statistic with $h=0$, which also provides good results.  The posterior for $\beta$ based on the simple summary is similar to that when the LNA summary statistic is used, but is slightly less precise for $\gamma$.

In terms of lazy ABC we use the value of $S(t)+I(t)$ at the end of the recording time as the lazy summary statistic.  In order to tune $h_{\mathrm{lazy}}$, we run ABC with the summary statistic formed from the LNA approximation only for a small number of iterations and recorded at each iteration the value of the simulated lazy summary statistic and whether or not the proposal was accepted or rejected.  From Figure \ref{figsub:summ_stat_lazy_acc} it is evident that most of the acceptances based on the actual summary statistic occur when the lazy summary statistic is between 70 and 110 (the observed value is 90).  Many simulations do not produce a lazy summary statistic within this range so that early rejection based on this lazy summary statistic seems like a good choice.  Therefore we set the lazy discrepancy to be the absolute value between the last observed and simulated data point and set $h_{\mathrm{lazy}} = 20$.  Note that if a proposal does not satisfy $h_{\mathrm{lazy}}$ then we continue nonetheless with probability $\alpha = 0.1$.  This seems like a reasonably conservative choice. 

The lazy ABC approach resulted in a very similar acceptance rate to usual ABC, 2.3\% and 2.5\% respectively, however the lazy ABC approach was about 3.5 times faster (roughly 16 hours down to 4.5 hours).

\subsection{Spatial Extremes Example} \label{subsec:spatial}

If it exists, the limiting distribution of the maximum of a suitably normalized
sequence of independent and identically distributed (multivariate)
random variables is in the family of multivariate extreme value distributions
(MEVDs). Max-stable processes are the infinite-dimensional generalization
of MEVDs.

Consider a set of extremal observations $y_{obs}^t = (y_{obs}^t(x_1),\ldots,y_{obs}^t(x_D)) \in \mathbb{R}^D$ collected at spatial locations $x_1,\ldots,x_D \in X \subset \mathbb{R}^p$ at time $t$. Here we define an extremal observation as the maximum of a sequence of independent and identically distributed random variables. Assume that $T$ independent and identically distributed extremal observations are taken  at each location, indexed by $t=1,\ldots,T$.  We denote the full dataset as $y_{obs}^{1:T}$ whereas all the data at the $i$th location is $y_{obs}^{1:T}(x_i)$.  It is possible to model such data using spatial models called max-stable processes (see, for example, \citet{Schlather2002} and \citet{Davison2012}).  The max-stable process arises from considering the limiting distribution of the maximum of a sequence of independent and identically distributed random variables.  Here we follow \citet{Erhardt2014} and focus on a max-stable process where each marginal (i.e.\ for a particular spatial location) has a unit-Fr\'echet distribution with cumulative distribution function $G(z) = \exp(-1/z)$.  Additional flexibility on each marginal can be introduced via a transformation with location ($\mu$), scale ($\sigma$) and shape ($\xi$) parameters.  Assuming that $Z$ has a unit-Fr\'echet distribution, the random variable 
\begin{align*}
Y &= \frac{\sigma}{\xi}(Z^\xi - 1) + \mu,
\end{align*}     
has a generalised extreme value distribution.  The first step, then, is to estimate the $(\mu,\sigma,\xi)$ parameters for each of the marginals separately based on the $T$ observations $y_{obs}^{1:T}(x_i)$ at each location, $i=1,\ldots,D$, to transform the data so that they, approximately, follow a unit-Fr\'echet distribution.  The data following this transformation we denote as $z_{obs}^{1:T}(x_i)$.  

For simplicity, we consider a realisation of this max-stable process at a particular time point, and thus drop the index $t$ for the moment.  Assume that the corresponding random variable for this realisation is denoted $Z = (Z(x_1),\ldots,Z(x_D))$.  Unfortunately the joint probability density function of $Y$ is difficult to evaluate for $D>2$.  However, an analytic expression is available for the bivariate cumulative distribution function of any two points, say $x_i$ and $x_j$ (with realisations $z_i$ and $z_j$), which depends on the distance between the two points, $h= ||x_i-x_j||$:
\begin{align}
G(z_i,z_j) = \exp\left(-\frac{1}{2} \left[\frac{1}{z_i} + \frac{1}{z_j} \right]\left[1 + \left\{1-2(\rho(h)+1)\frac{z_iz_j}{(z_i+z_j)^2}\right\}^{1/2} \right] \right), \label{eq:bivcdf}
\end{align}    
where $\rho(h)$ is the correlation of the underlying process.  For simplicity, we consider only the Whittle-Mat\'ern covariance structure
\begin{align*}
\rho(h) &= c_1\frac{2^{1-\nu}}{\Gamma(\nu)}\left(\frac{h}{c_2}\right)^\nu K_\nu\left(\frac{h}{c_2}\right),
\end{align*}
where $\Gamma(\cdot)$ is the gamma function and $K_\nu(\cdot)$ is the modified Bessel function of the second kind.  Note that there are several other options (see \citet{Davison2012}).  In the above equation, $0 \leq c_1 \leq 1$ is the sill, $c_2>0$ is the range and $\nu > 0$ is the smooth parameter.  The sill parameter is commonly set to $c_1 = 1$, which we adopt here.  Therefore interest is in the parameter $\theta = (c_2,\nu)$.  Here the prior on $\theta$ is set as uniform over $(0,20)\times(0,20)$.

A composite likelihood can be constructed \citep{Padoan2010,CooleyEtAl2011} since there is an analytic expression for the bivariate likelihood (i.e.\ the joint density of the response at two spatial locations), which can be obtained from the cumulative distribution function in \eqref{eq:bivcdf}.  The composite likelihood for one realisation of the max-stable process can be derived by considering the product of all possible (unordered) bivariate likelihoods (often referred to as the pairwise likelihood).  Then another product can be taken over the $T$ independent realisations of the process.  \citet{CooleyEtAl2011} utilise an adjusted composite likelihood directly within a Bayesian algorithm to obtain an approximate posterior distribution for the parameter of the correlation function.  We investigate a different approach and use the composite likelihood parameter estimate as a summary statistic for ABC.  The composite likelihood can be maximised using the function \emph{fitmaxstab} in the SpatialExtremes package in R \citep{Ribatet2013}.  For simplicity, we refer to this approach as ABC-cp (where cp denotes `composite parameter' to be consistent with \citet{Ruli2013}, who refer to their method as ABC-cs (`composite score')).

Our approach is compared with an ABC procedure in \citet{Erhardt2012}, who use a different summary statistic.  The method first involves computing the so-called tripletwise extremal coefficients.  One extremal coefficient calculation involves three spatial points.  \citet{Erhardt2014} use estimated tripletwise extremal coefficients, which for three spatial locations $i,j,k$ can be obtained using
\begin{align*}
\frac{1}{\sum_{t=1}^T 1/\max(z_{obs}^t(x_i),z_{obs}^t(x_j),z_{obs}^t(x_k))}.
\end{align*}
The full set of estimated tripletwise coefficients is high-dimensional, precisely $\binom{D}{3}$.    The dimension of this summary is reduced by placing the extremal coefficients into $K$ groups, which are selected by grouping similar triangles (formed by the three spatial locations) together.  This grouping depends only on the spatial locations and not the observed data.  \citet{Erhardt2012} then use the mean of the extremal coefficients within each group to form a $K$ dimensional summary statistic.  For the ABC discrepancy function, \citet{Erhardt2012} consider the L$_1$ norm between the observed and simulated $K$ group means.  For brevity, we refer to this ABC approach as ABC-ec where ec stands for `extremal coefficients'.  The reader is referred to \citet{Erhardt2012} for more detail.  This method is implemented with the assistance of the ABCExtremes R package \citep{Erhardt2013}.  

There are several issues associated with ABC-ec.  Firstly, it can be computationally intensive to determine the $K$ groups.  Secondly, there is no clear way to choose the value of $K$.  There is a trade-off between dimensionality and information loss, which may require investigation for each dataset analysed.  Thirdly, only the mean within each group is considered, whereas the variability of the extremal coefficients within each group may be informative too.  Finally, there is no obvious ABC discrepancy to apply.   In contrast, the ABC-cp offers a low dimensional summary statistic (same size as the parameter) and a natural way to compare summary statistics through the Mahalanobis distance (using an estimated covariance matrix of what is returned by \emph{fitmaxstab}).  However, the tripletwise extremal coefficients consider triples of locations (and so should carry more information compared with the pairwise approach of the composite likelihood) and also we find that computing the summary statistic of ABC-cp using \emph{fitmaxstab} is slower than a C implementation of the tripletwise extremal coefficients calculation (called into R using the Rcpp package \citep{Eddelbuettel2011}).  On the other hand, ABC-cp avoids the expensive clustering of triangles into groups.  

For both approaches, an MCMC ABC algorithm was used with proposal distributions carefully chosen to ensure a desired acceptance probability based on the results of some pilot runs.  ABC-ec was run for 2,000,000 iterations and the ABC tolerance chosen resulted in an acceptance rate of roughly 0.8\%.  Due to the extra computation associated with maximising the composite likelihood at each iteration, ABC-cp was run for 100,000 iterations with an ABC tolerance that results in an acceptance probability of roughly 8\%.  

Here we re-analyse the data considered in \citet{Erhardt2014}, which consists of the maximum annual summer temperature at 39 locations in midwestern United States of America between 1895 and 2009.  The data at each spatial location are firstly transformed to approximately unit-Fr\'echet margins by fitting a generalised extreme value distribution by maximum likelihood, and also taking into account a slight trend in the maximum summer temperatures over time (see \citet{Erhardt2014} for more details).  The max-stable process is then fitted to this transformed data using the ABC approaches described earlier. 

\begin{figure}
	\centering
	\subfigure[ABC-ec]{\includegraphics[height=0.3\textheight,width=0.45\textwidth]{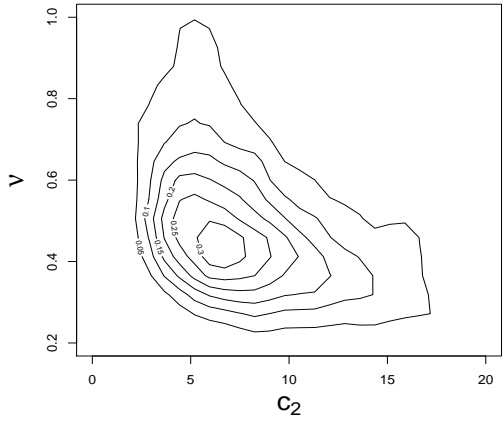}\label{figsub:ABCecContour}}
	\subfigure[ABC-cp]{\includegraphics[height=0.3\textheight,width=0.45\textwidth]{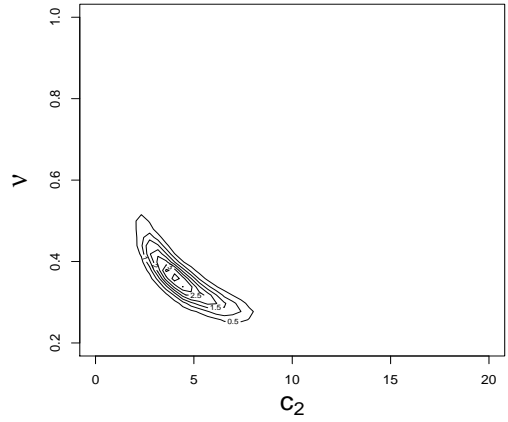}\label{figsub:ABCcpContour}}
	\subfigure[ABC-ec]{\includegraphics[height=0.3\textheight,width=0.45\textwidth]{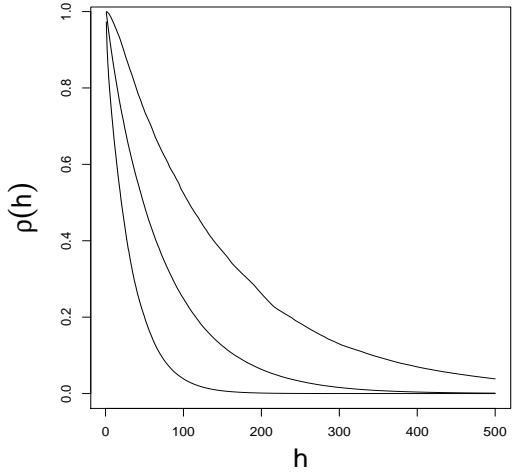}\label{figsub:ABCecRho}}
	\subfigure[ABC-cp]{\includegraphics[height=0.3\textheight,width=0.45\textwidth]{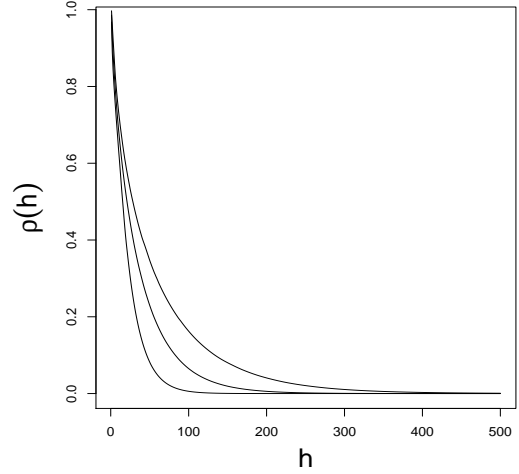}\label{figsub:ABCcpRho}}
	\caption{Posterior results for the spatial extremes example.  Shown are the bivariate posterior distributions of $(c_2,\nu)$ based on the (a) ABC-ec and (b) ABC-cp approaches.  The posterior median and 95\% credible interval of the spatial correlation function shown for the (c) ABC-ec and (d) ABC-cp methods.}
	\label{fig:SpatialExtremes}
\end{figure}

Contour plots of the bivariate posterior distributions for both the ABC-ec and ABC-cp approaches are shown in Figure \ref{fig:SpatialExtremes}.  Despite the much higher acceptance rate for ABC-cp, the resulting posterior for ABC-cp is substantially more concentrated compared with the results from ABC-ec.  Furthermore, it can be seen that the posterior spatial correlation function is determined much more precisely with ABC-cp.

Despite the encouraging results, a thorough simulation study is required to confirm the ABC-cp approach as a generally useful method for spatial extremes applications.  \citet{Erhardt2012} note that very different parameter configurations can lead to a similar correlation structure, as demonstrated by the `banana' shape target in Figure \ref{fig:SpatialExtremes}.  Therefore it may be inefficient to compare composite parameter summary statistics directly via the Mahalanobis distance.  Comparison through the composite likelihood itself may perform better.  Alternatively, if an expression for the composite score can be derived, then the ABC-cs method of \citet{Ruli2013} may be a computationally convenient approach for these sorts of models.  The ABC-cp method implemented here relies on already available functions in existing R packages.  Thus we leave the composite likelihood and score methods for further research.

\section{Discussion} \label{sec:Discussion}

In this chapter we provided a description of the indirect inference method and also detailed links between various likelihood-free Bayesian methods and indirect inference.  As highlighted by the examples in this chapter and other applications in articles such as \citet{Gallant2009,DrovandiEtAl2011,Gleim,DrovandiBII2014}, it is clear that the tools presented in this chapter can provide useful posterior approximations in complex modelling scenarios.

During this chapter we have also considered an extension of the reverse sampler of \citet{Forneron2016} using regression adjustment, an MCMC implementation of the lazy ABC approach of \citet{Prangle2014} and developed an ABC approach for spatial extremes models using the parameter estimate of a composite likelihood as the summary statistic.

A possible avenue for future research involves likelihood-free Bayesian model choice.  It is well-known that Bayes factors based on summary statistic likelihoods do not correspond to those based on the full data likelihoods \citep{Robert2011}.  It is typically not clear how to choose a summary statistic that is useful for model discrimination (although see, for example, \citet{Didelot2011,Estoup2012,Prangle2014a,Martin2014,Lee2014} for progress in this area); a summary statistic that is sufficient for the parameter of each model is still generally not sufficient for the model indicator (see \citet{Marin2013}).  An interesting direction for further research is to explore whether flexible auxiliary models can assist in developing likelihood-free methods that are useful for model selection in terms of Bayes factors and posterior model probabilities. 

\section*{Acknowledgements}

The author is grateful to Kerrie Mengersen for the helpful comments and suggestions on an earlier draft.   The author was supported by an Australian Research Council's Discovery Early Career Researcher Award funding scheme (DE160100741).  The author is an Associate Investigator of the Australian Centre of Excellence for Mathematical and Statistical Frontiers (ACEMS).

\bibliographystyle{apalike} 
\bibliography{refs}

\end{document}